\documentclass{iau} 
\usepackage{graphicx}
\usepackage{notoccite}
\usepackage[usernames,dvispsnames,svgnames]{xcolor}
\title[JD 11.~~The impact of CMEs on foF2] 
{The impact of CMEs on the critical frequency of F2-layer ionosphere (foF2)}
\author[Alene Seyoum, Nat Gopalswamy, Melessew Nigussie and Niguss Mezgebe]   
{Alene Seyoum$^{1,2}$, Nat Gopalswamy$^3$, Melessew Nigussie$^4$, and Nigusse Mezgebe$^{1}$
}\affiliation{$^1$Ethiopian Space Science and Technology Institute (ESSTI), Addis Ababa, Ethiopia. $^2$Dire Dawa University, College of Natural and Computational Science, Physics Department, Dire Dawa, Ethiopia. $^3$NASA Goddard Space Flight Center, Greenbelt, MD 20771, USA email: nat.gopalswamy@nasa.gov. $^4$Washera Geospace and Radar Science Laboratory, College of Science, Physics Department, Bahir Dar University, Bahir Dar, Ethiopia.}
\pubyear{2019}
\volume{356}  
\jname{Nuclear Activity in Galaxies Across Cosmic Time}
\editors{M. Povic, J. Masegosa, H. Netzer, P. Marziani, \\P. Shastri, 
	S. B. Tessema \& S. H. Negu,  eds.}
\begin{document}
	\maketitle
	\begin{abstract} The ionospheric critical frequency (foF2) from ionosonde measurements at geographic high, middle, and low latitudes are analyzed with the occurrence of coronal mass ejections (CMEs) in long term variability of the solar cycles. We observed trends of monthly maximum foF2 values and monthly averaged values of CME parameters such as speed, angular width, mass, and kinetic energy with respect to time. The impact of CMEs on foF2 is very high at high latitudes and low at low latitudes. The time series for monthly maximum foF2 and monthly-averaged CME speed are moderately correlated at high and middle latitudes.  \end{abstract}\firstsection 
\section{Introduction} Active regions on the Sun contribute to the variability of Earth's ionosphere, in particular to the variability of neutral and ionized densities. The ionosphere becomes variable due to lower atmospheric internal waves, and geomagnetic and  solar activity variations from the above atmosphere (Yi{\u{g}}it et al., 2016). The origin of space weather effects such as intense geomagnetic storms is due to CMEs (Gopalswamy, 2009). CME parameters are thought to cause a large volume of the Earth's ionosphere to increase ionization (Farid et al., 2015). In the ionospheric dynamo, equatorial electrojet is produced due to Eastward electric field (Seba and Nigussie, 2016). The electric field \textbf{E} interacts with Earth's magnetic field \textbf{B} causing strong vertical upward \textbf{E$\times$B} drift velocity and enhanced foF2 values in low latitudes (Horvath and Essex, 2003). This paper shows the impact of CMEs on foF2 from 1996 to 2018 in geographical high, middle, and low latitudes. This impact has importance for communication depending on relationship with solar activity. The parameters used in this work are monthly maximum value of foF2 and monthly-averaged values of CMEs angular width, speed, mass, and kinetic energy.
 
	\section{Data Sources}The foF2 data are obtained from ionosonde of UK Solar System Data Center CEDA-UKSSDC for three station data. These stations are SO166 ($67.400^{o}$N, $26.600^o$E) in Finland, BP440 (40.080$^o$N, 116.260$^o$E), in China, and VA50L (-2.700$^o$S, 141.300$^o$E) in Papua New Guinea\footnote{($\mathrm{https://www.ukssdc.ac.uk/wdcc1/ionosondes/secure/ion\_data.shtml}$)}. 
	The CMEs data are found from the SOHO/LASCO \footnote{($\mathrm{http://cdaw.gsfc.nasa.gov/CME\_list/}$)} (\cite{yashiro2004catalog}, \cite{gopalswamy2009soho}) in $23^{rd}$ and $24^{th}$ solar cycles from 1996 to 2018.  

	\section{Results} In Figure 1 we showed that the monthly maximum foF2 (right panels) tracks the monthly-averaged CME parameters (left panels) in 23$^{rd}$ and 24$^{th}$ solar cycles. In the ascending periods from 1997 to 2001, the maximum foF2 values are higher at low latitudes. In Figure 2 we showed the time series of CME speed and foF2. High and middle latitudes foF2 show similar time series to CME speed, but not for low latitudes. The foF2 at low latitudes (equatorial region) is rapidly fluctuating foF2 values, which needs further investigation. 
	
	\begin{figure} [h!]
		\begin{center}
			\includegraphics[width=0.4\textwidth]{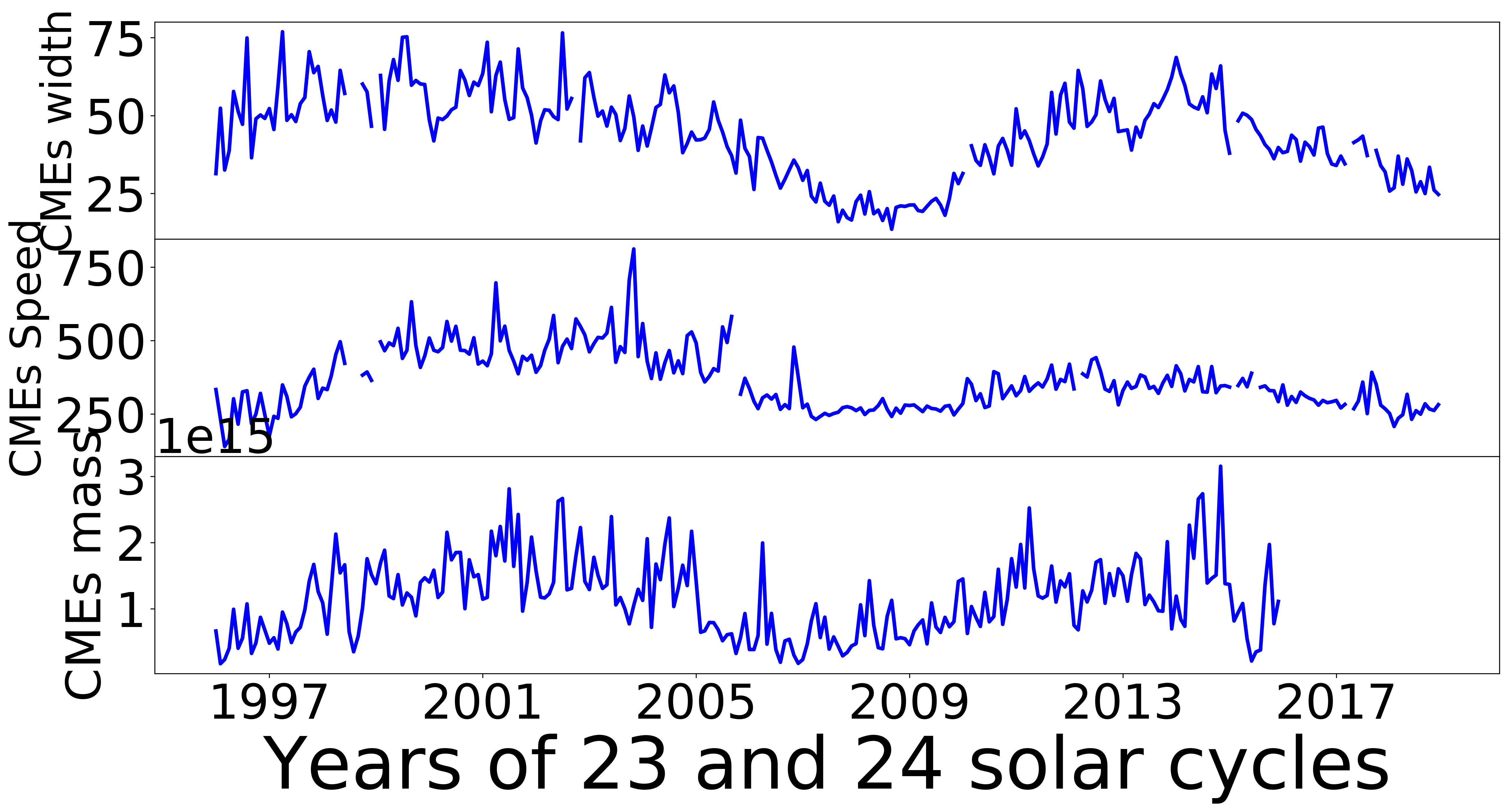}
			\includegraphics[width=0.38\textwidth]{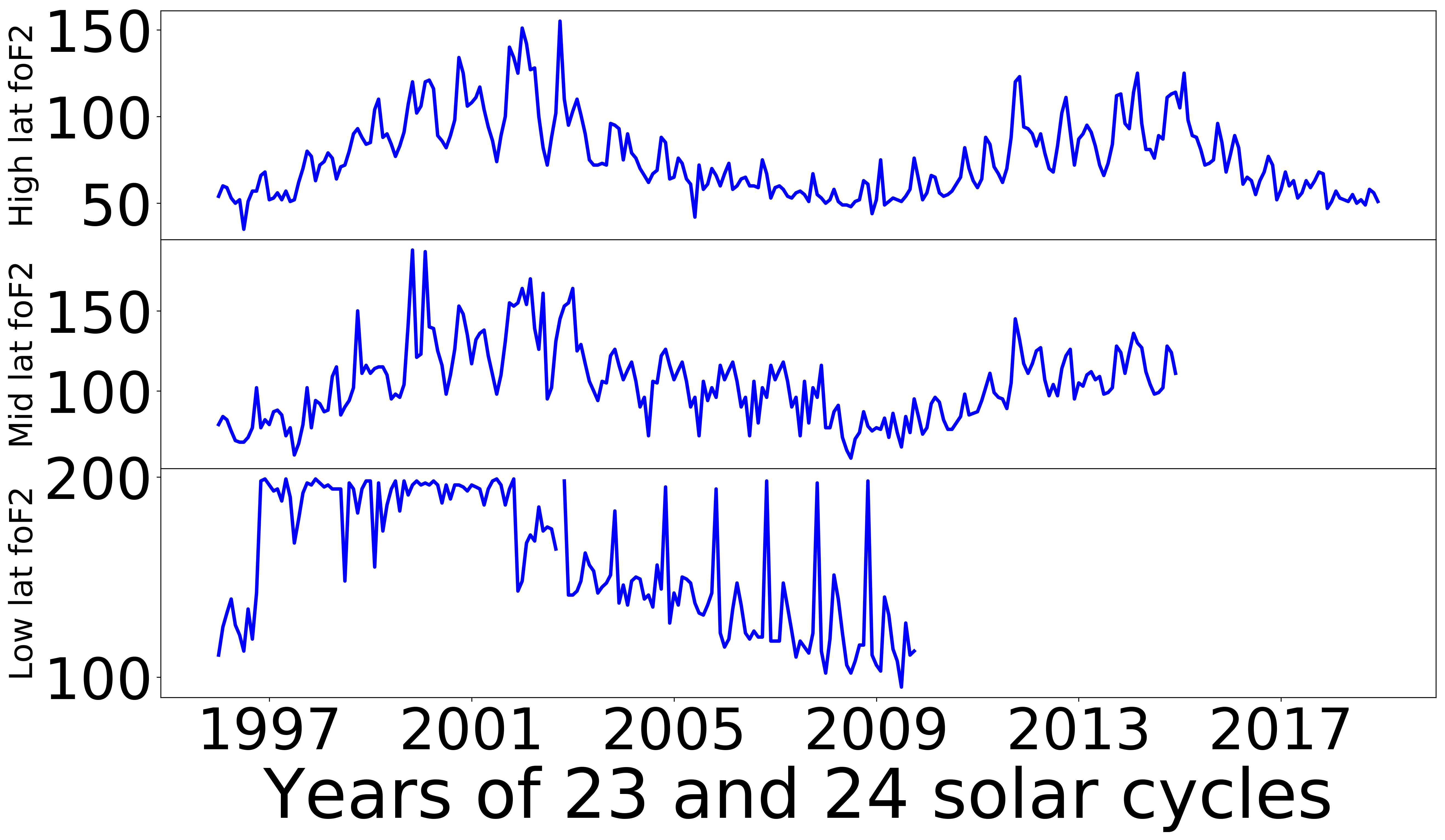}	
			\caption{\label{fig:}The relation of monthly-averaged CME parameters such as CME width (deg), CME speed (km/s), and CME mass (ergs) (left panels, from top to bottom) with monthly maximum foF2 values (in MHz) (right panels) from 1996 to 2018.}
		\end{center}
	
		\begin{center}
			\includegraphics[width=0.28\textwidth]{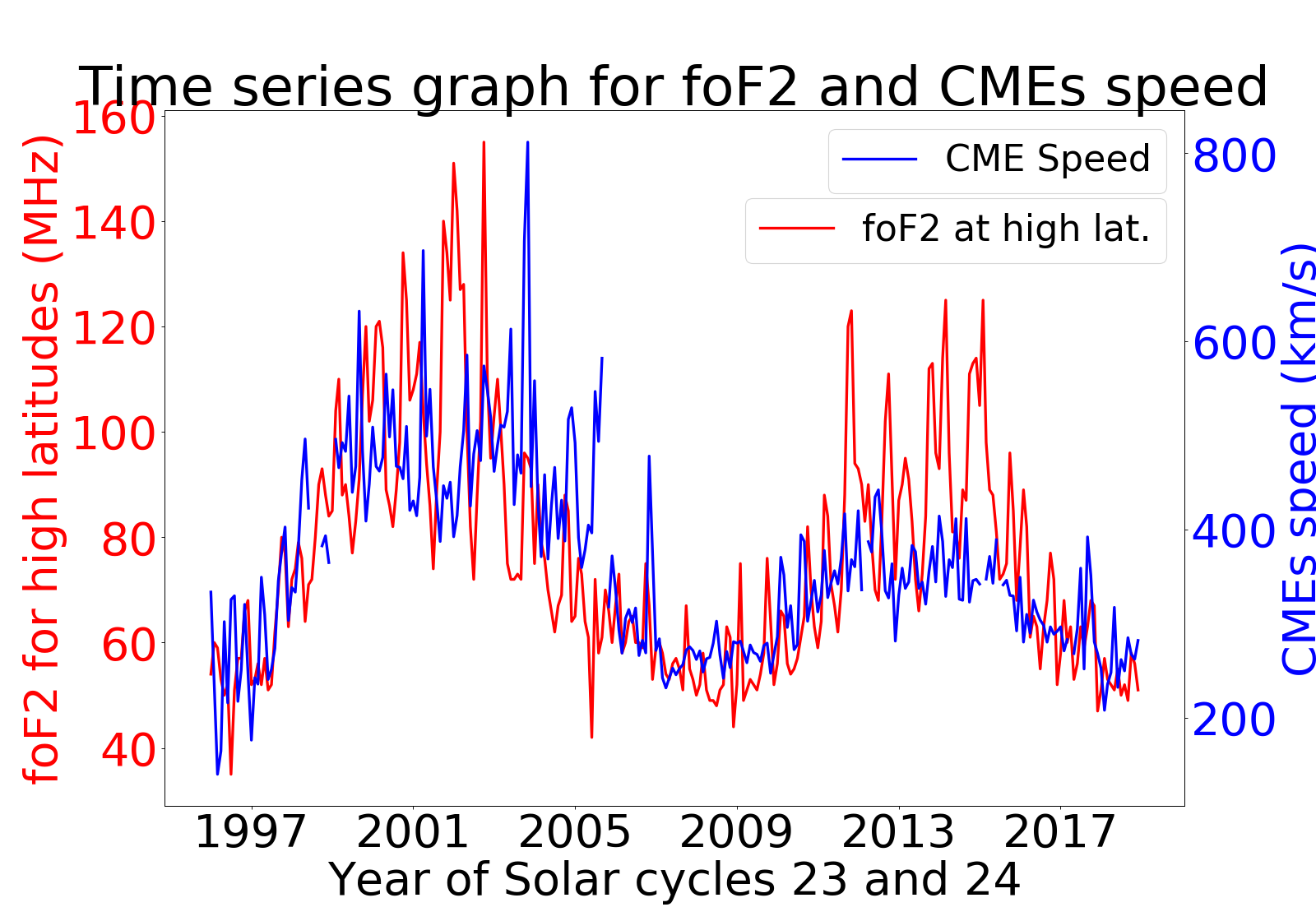}
			\includegraphics[width=0.28\textwidth]{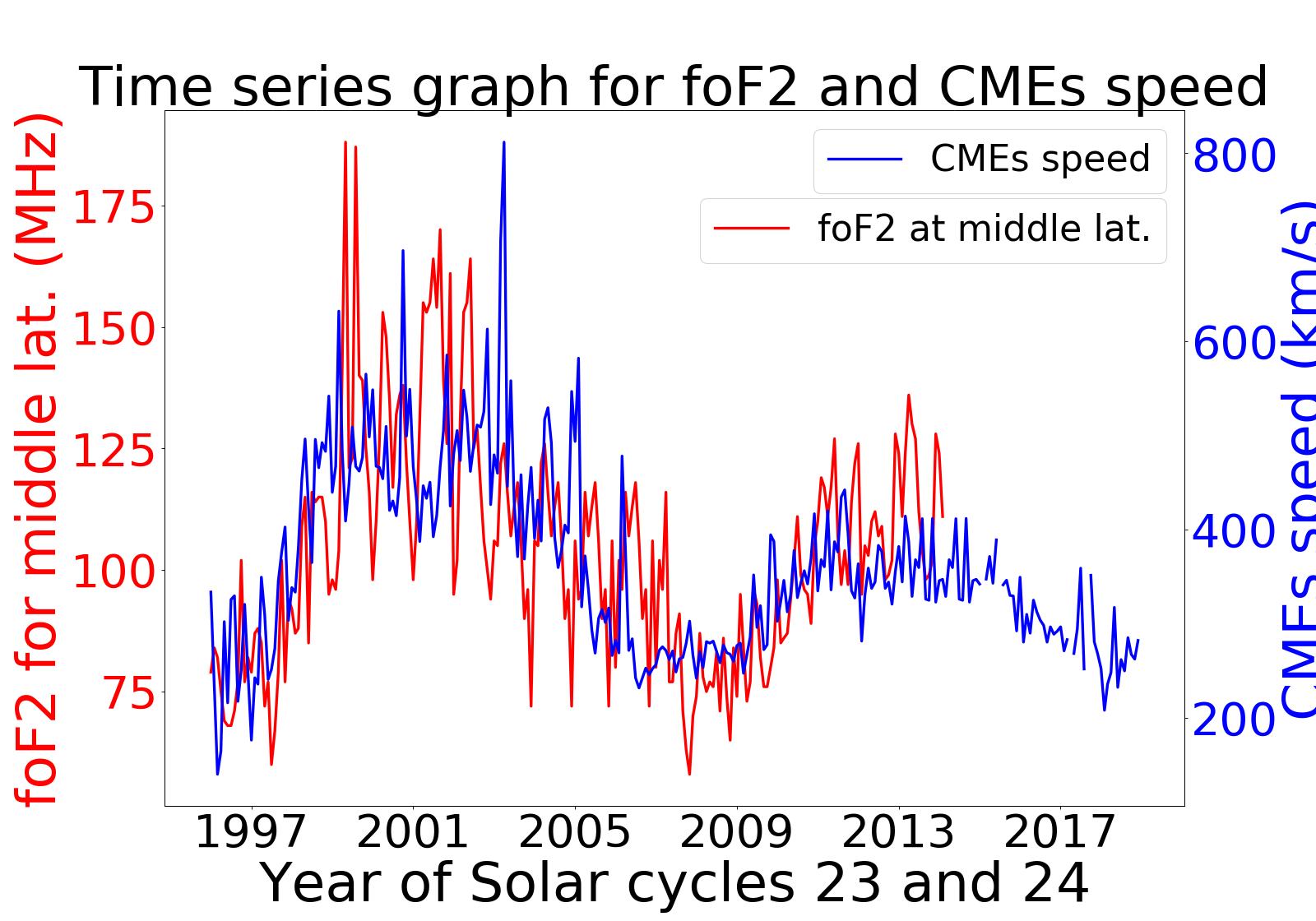}
			\includegraphics[width=0.28\textwidth]{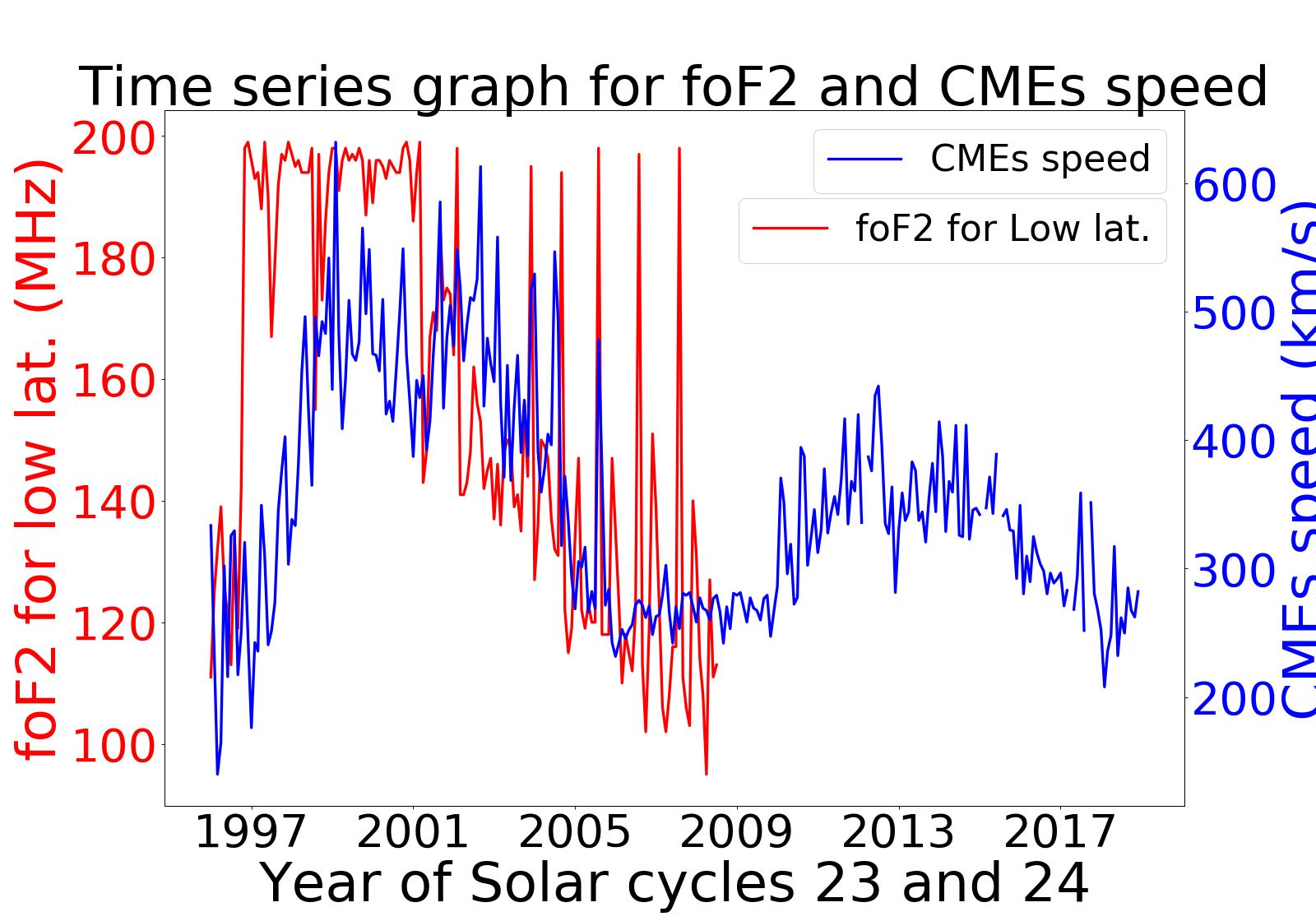}	
			\caption{\label{fig:}The time series of monthly-averaged CME speed in solar cycles of 23$^{rd}$ and 24$^{th}$ and monthly maximum values of foF2.}
		\end{center}
	
		\section{Conclusion} We studied the relation between monthly-averaged values of CME parameters and the monthly maximum value of foF2 at high, middle, and low latitudes. While there is moderately good correlation between the two time series graph at high and middle latitudes, for the low-latitudes of ionosphere correlation is not clear. Because the foF2 values here are higher and have fast fluctuation at low latitudes and it is not symmetric to  CME parameters at high and middle latitudes. This may be due to the \textbf{E$\times$B} upward drift velocity of the ionospheres at equatorial region (low latitudes) that can not be easily compressed and highly influenced by CMEs at high latitudes. This indicates the impact of CMEs on foF2 is higher at high latitudes (polar region) than at the low latitudes. The coincidence relation between maximum value of foF2 and CME parameters are important for modeling of space weather prediction at high and middle latitudes.
	\end{figure}


\begin{thebibliography}{}
		\bibitem[Hussein et al., 2015]{farid2015impacts}{Farid, H. M., Mawad, R., Yousef, M., and Yousef, S.} 2015, \textit{Elixir Space Sci}, 80
		\bibitem[Gopalswamy (2009)]{gopalswamy2009cme}{Gopalswamy N.,} 2009,
		\textit{Proceedings of IAU}, 5.S264
		\bibitem[Gopalswamy et al., 2009]{gopalswamy2009soho}{Gopalswamy N., Yashiro S., Michalek G., et al.,} 2009, \textit{  Earth, Moon, and Planets}, 104, 1-4
		\bibitem[Horvath and Essex, 2003]{horvath2003vertical}{Horvath, I., and Essex, E. A.} 2003, \textit{Annales Geophysicae}, 21, 4
		\bibitem[Seba and Nigussie, 2016]{seba2016investigating}{Seba, E. B., and Nigussie, M.} 2016,	\textit{Advances in Space Res.}, 58, 9
		\bibitem[Yashiro et al., 2004]{yashiro2004catalog}{Yashiro, S., Gopalswamy, N., Michalek, G., et al.,} 2004, \textit{JGR}, 109, A7
		\bibitem[Yi{\u{g}}it et al., 2016]{yiugit2016review}{Yi{\u{g}}it E., Kn{\'\i}{\v{z}}ov{\'a}, P. K., Georgieva, K. and Ward, W} 2016, \textit{JASTP}, 141, 1
		
	\end{thebibliography}
\end{document}